# NLOS Mitigation Using Sparsity Feature And Iterative Methods

Abolfathi. A, Behnia F. Marvasti F.
Department of Electrical Engineering
Sharif University of Technology
Email: abbas abolfathi@yahoo.com, behnia@sharif.edu, fmarvasti@sharif.edu

*Abstract*—well-known methods are employed to localize mobile station (MS) using line of sight (LOS) measurements. These methods may result in large error if they are fed with non LOS (NLOS) measurements. Our proposed algorithm, referred to as Sparse Recovery of NLOS using IMAT (SRNI), considers NLOS as unknown variables and solves the resultant underdetermined system emphasizing on its sparsity feature based on IMAT methods. Simulations are conducted to investigate the performance of SRNI in comparison of other conventional algorithms. Results demonstrate that SRNI is fast enough to deal with large combination of BSs and also accurate in lower number of BSs.

## I. INTRODUCTION

Many applications of next generation wireless communication systems rely on location estimation of mobile station (MS). It can be useful in network resource management, E-911 and hand off assistance [6]. However MS localizing adventures is not limited to communication systems. Psychologists are searching for the precise source location of an electric impulse in human brain [4]. These are the reasons of ongoing research on source location estimation.

Location estimation in different fields uses the same principles. In three dimensional space, Distance between transmitter and receiver defines a sphere centering each receiver. Intersection of these spheres leads us to the transmitter location. In an N dimensional space, at least N+1 receiver is needed to determine the location uniquely.

Time of arrival (TOA), time difference of arrival (TDOA) and signal strength measurement (RSS) are some of the well known techniques to measure the range between transmitter and receivers [1].

In the presence of noise, spheres do not intersect in one point, so it is necessary to find the most probable point to fit the source location. Some algorithms such as least square (LS) are developed to estimate this point [5]. LS assumes signals have traveled through line of sight (LOS) path. If one or more receivers see the transmitter through a reflector, measured range exceeds its real value. Such a non line of sight (NLOS) view caused large error in location estimation. This often happens in an urban environment.

Generally there are three main methods to cope with the NLOS measurements. In first method, by using the propagation characteristics of the channel and scattering model, MS location is determined. This method needs prior information about localization environment which is hard to achieve and varying time to time.

The second method attempts to use all LOS and NLOS measurements with considering a weighting to minimize effects of NLOS conditions. Residual weighting algorithm (RWGH) [1] find source location using all possible combination of receivers to calculate these weights. Unlike its accuracy, RWGH is time wasting, spatially when the number of receivers increases. Some recent algorithm such as weighted least-square by bounding box (BWLS) [2][4] are faster but not reliable in less receiver number.

In practical situation most of the receivers have a LOS view to transmitter. Therefore we can consider NLOS error as a sparse vector. Third method uses sparsity feature of NLOS vector for identification and mitigation of NLOS. Compressive sensing methods such as iterative reweighted LS (IRLS) are used broadly for this purpose [6][7].

In this paper we propose a novel algorithm attempts to find NLOSs by solving an equation system which considers NLOSs as unknown variables in addition to unknown position of MS. Such an equation system is underdetermined and cannot be solve by traditional methods.

Iterative method with adaptive thresholding (IMAT) is an effective solution for compress sensing problem introduced by F.Marvasti [10]. Our proposed algorithm, SRNI solve its underdetermine system by using sparsity feature of NLOS vector and based on IMAT.

The rest of this paper is organized as follows: SRNI algorithm and its requirements are introduced in section II. Simulation results are given in section III. Finally section IV concludes this work.

## II. ALGORITHM

Proposed algorithm is based on estimating of MS location by diverse combinations of BS. It also uses sparse vector reconstructor to determined NLOS vector by an underdetermined system of nonlinear equation. Taylor series estimator and IMAT methods are employed to obtain MS location under least square (LS) criterion and reconstruct NLOS sparse vector, respectively as our algorithm backbones. Firstly Taylor series estimator and IMAT method are discussed and finally proposed algorithm will be introduced.

### A. Taylor Series Estimator

Suppose MS and BS are located in a two dimensional space. Let N BS be placed in arbitrary location around the MS. Position of ith BS can be expressed as bellow:

$$(x_i, y_i) \quad i = 1 \ldots N \quad (1)$$

Measured range by each BS ($r_i'$) is modeled as follow:

$$r_i' = \sqrt{(x_i - x_m)^2 + (y_i - y_m)^2} + n \quad (2)$$

Where $(x_m, y_m)$ represents Cartesian coordinates of MS at unknown position. n denotes measurement noise as zero-mean, independent and Gaussian distributed random process.

A non linear system consists of N equation and 2 variables which are MS unknown location is represented in (2). Usually N is greater than required and the system is cropped by noises. So a unique intersection for all equations is hard to find and it is needed to select the most probable answer as MS location.

LS tries to find the estimation that minimize the sums of error squares over all measurements. In mathematical form LS can be state as:

$$(x_m, y_m) = argmin_{x_m, y_m}(\sum |r_i' - r_i|^2) \quad (3)$$

Where $r_i$ is the accurate distance between ith BS and MS and it can be written as:

$$r_i(x_m, y_m) = \sqrt{(x_i - x_m)^2 + (y_i - y_m)^2} \quad (4)$$

An initial guess can be useful to solve nonlinear system in (2). Let $(x_v, y_v)$ be our initial guess for MS position. It can be written that:

$$x_m = x_v + \delta_x \qquad y_m = y_v + \delta_y \quad (5)$$

Where $\delta_x$ and $\delta_y$ denote difference between true MS position and our guess. Using Taylor series expansion of equations in (2) and considering only terms below second order we will have:

$$r_i' = \sqrt{(x_i - x_v)^2 + (y_i - y_v)^2} + a_{ix}\delta_x + a_{iy}\delta_y \quad (6)$$

Where:

$$a_{ix} = \frac{\delta r_i}{\delta x}\bigg|_{(x_v, y_v)} = \frac{x_v - x_i}{\sqrt{(x_i - x_m)^2 + (y_i - y_m)^2}}$$

$$a_{iy} = \frac{\delta r_i}{\delta y}\bigg|_{(x_v, y_v)} = \frac{y_v - y_i}{\sqrt{(x_i - x_m)^2 + (y_i - y_m)^2}}. \quad (7)$$

We can define following matrix to compact the equations:

$$A = \begin{bmatrix} a_{1x} & a_{1y} \\ a_{2x} & a_{2y} \\ \vdots & \vdots \\ a_{Nx} & a_{Ny} \end{bmatrix} \quad \delta = \begin{bmatrix} \delta_x \\ \delta_y \end{bmatrix} \quad z = \begin{bmatrix} r_1' - r_1(x_v, y_v) \\ r_2' - r_2(x_v, y_v) \\ \vdots \\ r_N' - r_N(x_v, y_v) \end{bmatrix}$$

$$e = \begin{bmatrix} e_1 \\ e_2 \\ \vdots \\ e_N \end{bmatrix}. \quad (8)$$

With these definitions, equations in (2) can be rewritten as matrix form as follows:

$$A\delta = z - e \quad (9)$$

$\delta$ which could minimize error squares in (3) can be expressed as:

$$\delta = [A^T R^{-1} A]^{-1} A^T R^{-1} z \quad (10)$$

Where **R** is a weighting matrix controlling convergence speed and can be chosen as covariance or identical matrix. By using iterative method and updating $x_v$ and $y_v$ as bellow, MS position satisfying LS will be achieved.

$$\begin{aligned} x_v &\leftarrow x_v + \delta_x \\ y_v &\leftarrow y_v + \delta_y \end{aligned} \quad (11)$$

Taylor series estimator is one of the most accurate methods to estimate location from range measurements, but its convergence is not guaranteed and relies on initial guess.

### B. Iterative Method with Adaptive Thresholding (IMAT)

A vector is known as sparse which most of its coefficients are zero in a domain. If there are more than 50 present non-zero coefficients, the vector is called dense [11].
Let **S** be $M \times 1$ vector which is sparse in $\Psi$ domain. Consider vector **Y** as our observation to determine **S**. Assume **Y** has L coefficients. $\phi$ is $L \times M$ measurement matrix relating **Y** and **S** as follows:

$$Y = \phi \times S \quad (12)$$

The problem is how to find **S** when only **Y** and $\phi$ are known.
If $L \geq M$ then **S** will be determined as:

$$S = (\phi^T \phi)^{-1} Y \quad (13)$$

If $L < M$ the problem could not be solved by traditional methods. However by knowing that S is sparse in $\psi$ domain, it can be determined uniquely even if L is less than M.
Transform matrix $\varphi$ is employed to represent **S** in its sparse domain as **X**.

$$X = \varphi \times S \quad (14)$$

IMAT is an iterative method which attempts to find S using its sparsity feature. IMAT retrieves sparse vector by alternating projections between the information domain and the sparse domain. IMAT thresholds sparse vector components in each iteration. It considers some of **S** components are known for sure. In this case, known components of **S** should be replaced in new **S** vector resulting by IMAT, in each iteration. IMAT also assumes **Y** is a masked form of **S**. it means measurement matrix just masks some components of **S**.

A simple pseudo-code is presented for IMAT algorithm bellow.

---

**Algorithm 1: Iterative Method with Adaptive Thresholding (IMAT)**

**Input:**
- **Y**:= observation vector which has been masked
- **M**:= mask vector

**Output:**
- **S**:= recovered vector

**Procedure IMAT(Y,M)**
1. $x^0 = Y$
2. For i = 0 : iter_max
3. $X^i = \varphi \times x$
4. $X^i_{threshold} = Threshold(X^i)$
5. $x^{i+1}_{new} = \varphi^H \times X^i_{threshold}$
6. $Where\ mask = 1, update\ x^{i+1}_{new} \coloneqq Y$
7. End for
8. $return\ S \coloneqq x^{iter\_max}_{new}$

**End procedure**

---

Note that when the location of the sparsity is unknown, at least L must be twice the sparsity number to recover **S** [12].

*C. Sparse Recovery of NLOS using IMAT (SRNI)*

Considering NLOS condition, range measurements $r'$ can be expressed as:

$$r'_i = \sqrt{(x_i - x_m)^2 + (y_i - y_m)^2} + n_i + NL_i \qquad (15)$$

Where $n_i$ is measurement noise same as which is used in (2). $NL_i$ denotes NLOS error for ith BS due to signal traveling through non straight path. According to triangle inequality law, NL is always non negative. NLOS vector can be defined as:

$$\mathbf{NL} = \{NL_1, NL_2, \dots, NL_N\} \qquad (16)$$

There are N equations and two MS location variables and N variables stands for NLOS. In addition because of non linearity of (15) one more measurement and totally N+3 equations are needed to solve the system uniquely. Therefore traditional methods are not able to address the problem.

Practical experiments indicate most of the BSs have LOS view to MS so **NL** vector is sparse. Using sparse property of **NL** vector IMAT can solve the problem. IMAT requires a transform to alternate measurements space $\{r'\}$ to sparse space $\{NL\}$.

NLOS error shifts MS estimated location from its true value. This replacement is backward the NLOS BS. For simplicity it is assumes that only first BS faces NLOS error among all N BSs. Localizing using a combination of BSs except first BS will result in true location. We call it test combination for first BS. The distance between estimated location obtaining by test combination and first BS $r''_1$ can be calculated by (4). $NL_1$ can be calculated as:

$$NL_1 = r'_1 - r''_1 \qquad (17)$$

Where $r'_1$ is the range measured by first BS.

In the case when there are more than one NLOS BS, localizing with test combination cannot result in accurate position and **NL** will be cropped by error. However it is still close enough to true value for IMAT. This condition is illustrated in simulation study.

Proposed transform can be concluded as following pseudo-code:

---

**Algorithm 2: Transform of Range Measurement to NLOS Space**

**Input:**
- $r'$:= Range measured by BSs
- $N$:= Number of all BSs
- $(x_i, y_i)$:= position of BSs

**Output:**
- **NL**:= NLOS vector

**Procedure Transform($r', N, (x_i, y_i)$)**
1. For i=1:N
2. $S' \coloneqq$ test combination by ignoring ith BS
3. $X'_m \coloneqq$ MS estimated position by LS and using $S'$
4. $r''_i = |(x_i, y_i) - X'_m|$
5. $NL_i = r'_i - r''_i$
6. End for
7. $return\ \mathbf{NL} = \{NL_1, NL_2, \dots, NL_N\}$

**End procedure**

---

To use IMAT, a thresholding method must be chosen. According to the fact the most shift in estimated BS location is backward the BS with the greatest NLOS error, we can only choose greater component of NL vector and let others be zero in thresholding sequence.

Now we are able to employ IMAT to solve (15) for finding NL vector and to estimate the MS position by modified measurements as SRNI algorithm. SRNI is presented in following pseudo-code:

---

**Algorithm 3: Sparse Recovery of NLOS using IMAT (SRNI)**

**Input:**
- $r'$:= Range measured by BSs
- $N$:= Number of all BSs
- $(x_i, y_i)$:= position of BSs

**Output:**
- $X_m'$:= estimated position of MS
- $M$:= Number of NLOS BS identified by SRNI
- **NL**:= estimated NLOS vector

**Procedure SRNI($r', N, (x_i, y_i)$)**

1. $NL \coloneqq zeros(1, N)$
2. For i = 0 : iter_max
3.     **NL**:=Transform$(r' - NL, N, (x_i, y_i))$
4.     **Threshold_NL**:=Threshold **NL** by passing largest component in magnitude and letting others be zero
5.     **NL**=**NL**+ **Threshold_NL**;
6. End for
8. $X_m'$: MS estimated position by LS considering $(r' - NL)$ as range measurement.
7. M:Number of NL component greater than a certain value
8. $return\ X_m', M, NL$

**End procedure**

Note that in 7th line of SRNI pseudo-code a certain value for thresholding is needed than can be considered as standard deviation of noise measurement and also in 4th line large component is chosen by absolute value but return in signed value.

Let M and N be the number of BS with NLOS error and the number of all BS respectively. In two dimensional space three measurements is required to localizing so there are N-3 extra measurements. IMAT needs equations at least twice the M to recover NL vector. Therefore M should satisfy following condition to empower SRNI:

$$M < \frac{N-3}{2} \quad (18)$$

Using this limit we can define a valid zone for SRNI. If the number of NLOS BS reported by SRNI is more than M, results are unreliable.

### III. SIMULATIONS

In this section simulation results are presented which are conducted to investigate the performance of the proposed algorithm. Six different scenarios are developed to study SRNI in whole aspects. For comparison some other well-known algorithm such as LS [13], BB [4], BWLS [2] and RWGH [1] are applied too. However it is not expected that LS suppresses the NLOS error but it is studied as the backbone of SRNI.

Adaptive white Gaussian noise model is employed to simulate noise in range measurement. For simplicity all BSs are assumed to have same standard deviation for noise measurement. Standard deviation of noise measurement is considered to be fixed to 60m unless otherwise stated. To be free of noise behavior, 1000 independent trials are used in each simulation. In all scenarios BSs number 1 to 8 are located at (6000,0), (3000,-6000), (-3000,-5000), (-6000,-1000), (-4000,6000), (0,5000), (4000,6000), (-6000,4000) respectively in meters. Also MS is at (2000,1000). Fig. 1 illustrates the geometry.

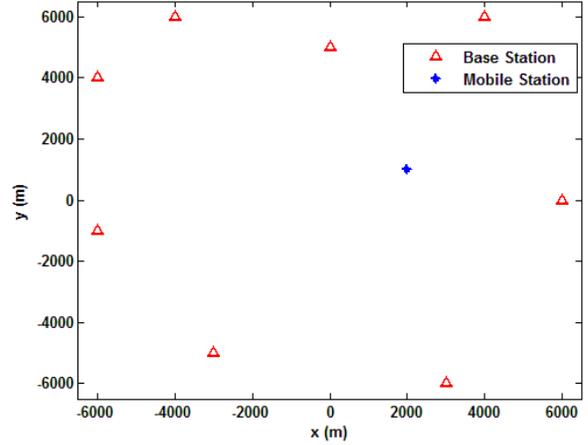

Fig.1. localization geometry

NLOS error vector for BSs is considered as follow in meters:

$$NL = [1000, 500, 800, 750, 400, 0, 0, 0] \quad (19)$$

But it may we use only some of these NLOS errors. It is clearly stated which BSs are dealt with NLOS in each scenario description.

To evaluate different results, root mean square error (RMSE) of estimated position is used which is calculated as:

$$RMSE = \sqrt{\frac{\sum_{i=1}^{1000}(\hat{x}_{mi}-x_m)^2+(\hat{y}_{mi}-y_m)^2}{1000}} \quad (20)$$

Where $(\hat{x}_{mi}, \hat{y}_{mi})$ and $(x_m, y_m)$ are the estimated and true position of MS in ith experiment.

Initial guess of LS and SRNI algorithm consider being MS true position to provide assurance that these algorithm will converge.

#### A. Iteration number response

This section attempts to show the affection of iteration number of SRNI on its performance. It is supposed only first two BSs are cropped by NLOS. Fig. 2 depicts the results.

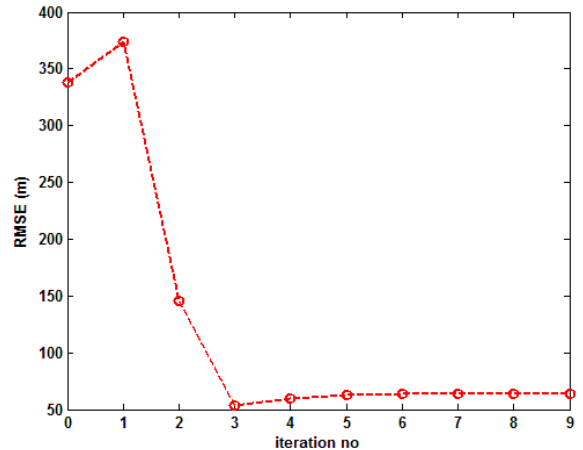

Fig.2. SRNI performance with different Iteration No.

As it is appears after 3 iterations SRNI converge to its final state. Also it remains stable after extra iterations. Iteration number of 10 is selected as a sufficient and trusted value in the rest of simulations.

### B. Resistance against noise

To evaluate the performance of different algorithm dealing with noisy condition, we vary standard deviation of noise from 0 to 100 meters. NLOS is assumed only for first BS. Fig. 3 presents the results.

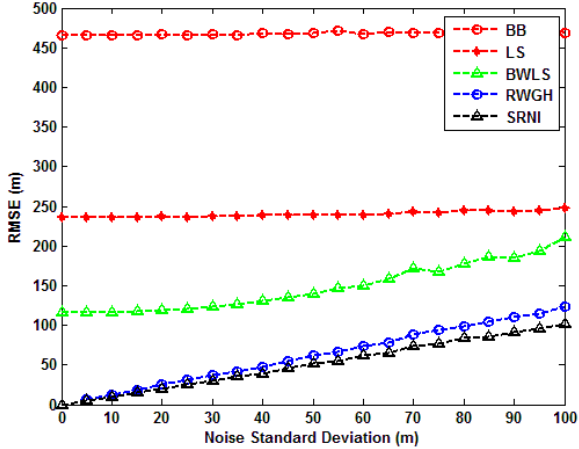

Fig.3. Influence of noise on localizing methods performance

This simulation indicates that RWGH and SRNI have a comparable RMSE to noise standard deviation however as noise increases SRNI perform better than RWGH. In noiseless condition RWGH and SRNI will result in exact position.

### C. Resistance against NLOS error

In this part ability of algorithms to suppress NLOSS error is studied. It is considered that only first BS is cropped by NLOS which varies from 0 to 1000 meter. Fig 4 illustrates the results.

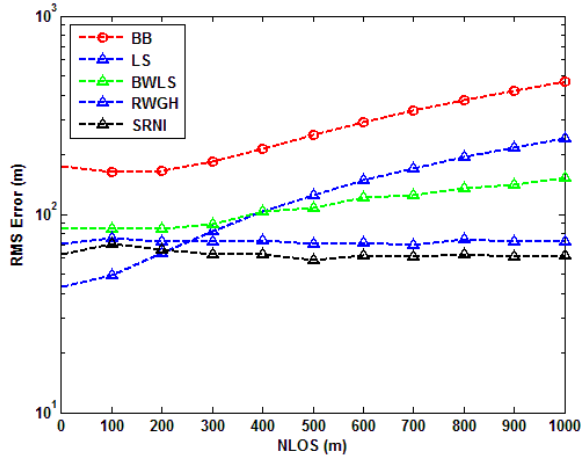

Fig.4. Influence of NLOS increment on localizing methods performance

LS and BB cannot mitigate the NLOS error and this is visible in this figure as the main weakness of BB and LS.

BB is less affected by NLOS rather than LS. In other hand, LS is the best algorithm in LOS condition. RWGH and SRNI suppress the error in all over the NLOS variation however SRNI performs better than RWGH.

### D. Increscent in NLOS station number

To investigate the performance of algorithms against NLOS station number, this simulation is conducted. According to inequality (18) SRNI can suppress the error up to 2 number of NLOS station due to the fact there are 8 stations in general. To examine the condition with m number of NLOS station, it is assumed that the first m stations have NLOS according to (19). Fig. 5 confirms the inequality of (18).

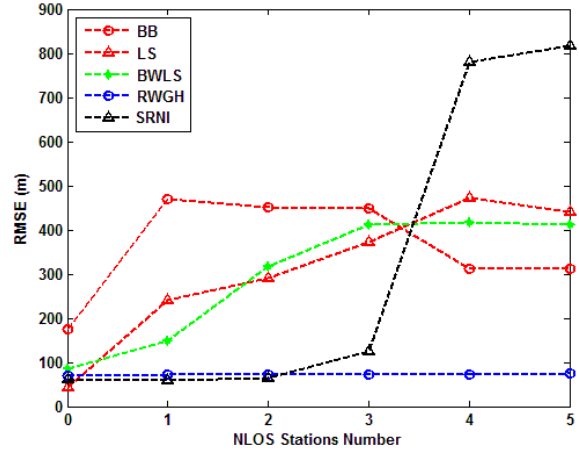

Fig.5. Influence of increment in NLOS number on localizing methods performance

### E. Increment in total station number

This subsection reveals the performance of algorithm against increment in total station number. Only first BS experiences NLOS. It is expected greater number of station results in better localizing. However it is not true in case of RWGH. LS performance improves which is the main idea in generation of proposed transform. BWLS is unable to estimate position correctly when there are a few stations. SRNI and RWGH have acceptable performance in all conditions. Results are presented in Fig. 6.

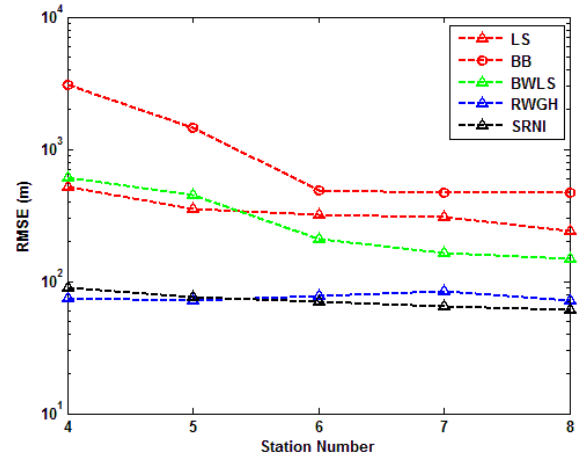

Fig.6. Iinfluence of increment in total BS number on localizing methods performance

### F. Time efficiency

This simulation indicates the time efficiency of algorithms against the increment of the stations. Results are reported for an Intel corei5 2.4 GHz CPU. Fig 7 depicts the time performance of different algorithm.

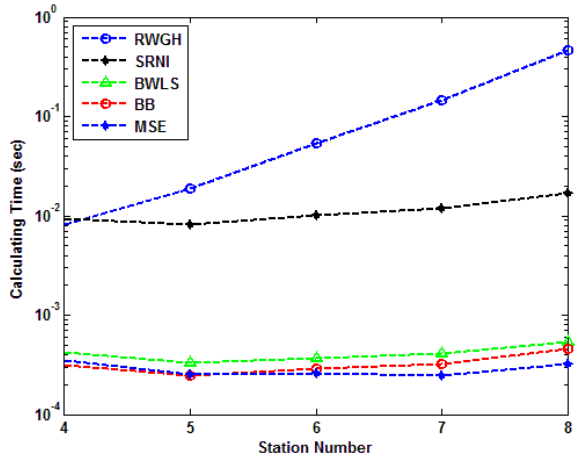

Fig.7. Timing performance of different methods

As it is appear RWGH requires calculating time in an exponential manner related to station number. So it is not possible to use RGWG in large combination of BSs. SRNI calculating time is linear to station number. Therefore SRNI is still fast enough when there are numerous BSs.

### IV. CONCLUSION

Simulation results indicate that large error in estimated location of MS can be experienced if NLOS is not suppressed. We introduced an extra variable standing for NLOS for each BS to determined NLOS value of BS. Therefore the number of variables is more than equations. Proposed algorithm, SRNI, employs IMAT to solve this underdetermined system. The only required consideration is sparsity of NLOS vector. The maximum NLOS number that SRNI can deal with is calculated. This limit can be used as a trust factor for SRNI results. The comparison of different conventional algorithms is presented by simulation study. It is shown that SRNI has the better performance among all algorithms in its valid zone. SRNI unlike RWGH is fast enough to be employed for large number of BS. It can be stated that SRNI performs as the best algorithm in large combination of BSs and still remains accurate in lower number of BSs.